\documentclass[12pt]{article}
\begin{document}

\title{Severe Vesico-ureteral Reflux and Urine Sequestration:
Mathematical Relations and Urodynamic Consequences.}

\author{Lisieux Eyer de Jesus$^1$\thanks{lisieux@uol.com.br}, \,
Paulo de Faria Borges$^2$\thanks{pborges@cefet-rj.br}
\\
\\
\small \it
$^{1}$Pediatric Surgery Department, 
Hospital Universit\'ario Antonio Pedro\\ 
\small \it Av. Marques de Paran\'a, Centro, 
 \\
\small \it 20000-000 Niter\'oi, Brazil\\
\\
\small \it
$^{1}$Hospital dos Servidores do estado, 
Brazilian Ministry of Health\\ 
\small \it Rio de Janeiro, Brazil
\\  
\\ 
\small \it
$^{2}$Centro Federal de Educa\c c\~ao 
Tecnol\'ogica Celso Suckow da Fonseca\\
\small \it Coordena\c c\~ao de F\'{\i}sica,  
Av. Maracan\~a, 229, Maracan\~a, \\
\small \it 20271-110 Rio de Janeiro, Brazil}

\bigskip

\maketitle

\pagebreak

\begin{abstract}

Some simple mathematical formulae to calculate the volumes of proximal
pyeloureteral reflexive systems are presented, and the results are compared to 
bladder capacity values. Using the results of the calculi, the author discusses 
possible implications of severe urinary sequestration in the pyeloureteral systems.
Using geometrical and topological approximations we calculate 
the volumes of ureters and renal pelvises, applying in vivo  measurements obtained 
from conventional ultrasound, retrograde cystourethrograms and topographic anatomic 
references. Approximations use 2 decimals and assumed $\pi$ value was 3.14.
Ureteral and pyelic volumes are calculated, respectively, from the mathematical 
formula for the cylinder and cone volumes. Dolicomegaureter are compensated using 
proportional calculi. Bladder volumes are estimated from conventional formulae.
Proximal urinary sequestration is compared between infants and older children with VUR.
Mechanisms of direct induction of bladder urodynamic failure from VUR are suggested.
Sequestration of urine in the ureter and renal pelvis can be estimated from 
mathematical formulae in patients with VUR. The values used derive from ultrasound 
examinations, CUM and topographical anatomical references. Primary VUR can determine 
urodynamic problems. Urine sequestration in the proximal urinary system is worse in 
infants than in older children.

KEY-WORDS:
Vesico-ureteral reflux – urodynamics – infants – bladder failure  

\end{abstract}

\pagebreak

{\centering{LEGENDS:}}

\noindent VUR: vesicoureteral reflux
 
\noindent CUD: conventional urodynamic 

\noindent VUD: videourodynamics 

\noindent DUS: Dynamic ultrasound 

\noindent US: conventional ultra-sound

\noindent CUM:  micturating cystourethrographies

\noindent AP: antero-posterior

\noindent CT: computerized tomography

\noindent MRI: Magnetic resonance imaging

\noindent pv: pelvis volume

\noindent uv: ureteral volume

\noindent B: major cone axis

\noindent b: minor cone axis

\noindent Rul: ureteral length from CUM 

\noindent Rptd: Real pelvis-trigone distance from CUM

\pagebreak

\section{INTRODUCTION}

Severe vesicoureteral reflux (VUR) renders the interpretation of conventional 
urodynamic (CUD) tests results difficult. Urine sequestrated in the ureter and 
renal pelvis may be summed to bladder urine and simulate a normal or even 
augmented bladder capacity in low volume / low compliance bladders. This phenomenon 
can induce serious doubts to indicate or not a bladder augmentation in dysfunctional 
or neuropathic bladder patients. If the clinical question is "Should we augment 
this patient because of his low cystometric capacity?" and the patient demonstrates 
a severe dilating VUR CUD may not give the correct answer.
The same biophysical problem applies to analyze the frequent association between 
severe VUR and bladder dysfunction, especially in infants. In those children an 
empty bladder after a normal complete micturition may be immediately followed by 
bladder repletion with urine from the full ureters/ renal pelvises, depending on 
the proportion of sequestered volume in the high urinary tract to the bladder 
capacity and on the clearance of the proximal urinary tract. This causes high 
urination frequency, high bladder pressure, urinary stasis and/or bladder urodynamic 
decompensation, depending on the case. In patients who depend on periodic urethral 
catheterizations to empty the bladder (CIC) the immediate bladder re-filling after 
micturition determines persistent bladder stasis and its consequences 
(urinary tract infection – UTI and urinary incontinence).
The solution to this dilemma has been to substitute CUD for videourodynamics (VUD). 
With VUD the operator can confront bladder pressures or volumes with timely images 
obtained with fluoroscopy. This allows him/her to define the exact bladder capacity 
immediately before any reflux to the ureter ("real" functional bladder capacity). 
Nevertheless, VUD is technically difficult to obtain (especially in children), VUR 
is a dynamic phenomenon (VUD demonstration can vary between different bladder filling cicles) 
and may occur in extremely low pressures and volumes. VUD is also costly, needs 
specific expertise and involves relatively high radiation doses 
(to the patient and to the equipment operator). Active movements and the position 
of the patient influence the results. Dynamic ultrasound (DUS), newly described \cite{fil} , 
is also an option, with the advantages of an easy availability of the equipment, no 
irradiation and a more "physiologic" exam. The disadvantages are the high expertise 
involved, long examining times, need of cooperation from the patient, recent 
description, limited experience with the method, high cost and variations of results 
between different examiners. 
A new answer to this problem would be to calculate the volumetric capacity of the full 
pyeloureteral system in each patient, and this is the prime objective of this article. 
Some easy formulae to calculate those volumes are presented, independently from urodynamic 
tests. Using this theoretical model, the author discusses possible implications of 
severe urinary sequestration in the pyeloureteral systems.

\section{MATERIALS AND METHODS:}

Using geometrical and topological approximations we calculate the volumes of ureters 
and renal pelvises, applying 'in vivo' measurements obtained from conventional 
ultra-sound (US) and cystourethrographies (CUM). Approximations use 2 decimals and 
assumed $\pi$ value was 3.14.

\section{RESULTS:}

\subsection{Preliminary data:}

The renal pelvis is (geometrically) an elliptical cone. We can calculate an elliptical 
cone volume by the formula:

\begin{equation}
pv={\pi h.B.b \over 3}
\end{equation}

\noindent Considering $\pi = 3.14$,

\begin{equation}
pv={3.14\times h.B.b \over 3}
\end{equation}

\noindent And 

\begin{equation}
pv={1.04\times h.B.b}
\end{equation}

The highness and the base diameters of the renal pelvis are obtained by ultrasound, 
with a full bladder. Using this formula the pelvic volume is 1.04 multiplied by the 
3 dimensions measured (highness, depth and AP diameter of the pelvis). It is absolutely 
necessary that the radiologist obtains those dimensions with a full bladder, to 
replicate the physiologic moment of a micturition/ bladder emptying and to consider 
diameter measurements from the pyeloureteral junction to as close as possible to the 
renal sinus (so as not to consider a shorter pelvis than real and obtain false low volumes).

The ureter is, geometrically, a cylinder. The formula to calculate a cylinder volume is:

\begin{equation}
uv=\pi r^2 h
\end{equation}

\noindent Considering $\pi=3.14$ and $radius(r)=diameter/2$,

\begin{equation}
uv= 3.14\times\left({diameter \over 2}\right)^{2}h
\end{equation}

\noindent Using Palmer's approximation \cite{palmer} , ureteral length may be 
approximated from the formula $age + 10(cm)$,

\begin{equation}
uv= 3.14\times\left({diameter \over 2}\right)^{2}\times(age(years)+10)
\end{equation}

Alternatively, ureteral length (pyeloureteral junction / uretero-vesical junction) 
can be estimated in vivo, using classical concepts from topographic anatomy \cite{testut} , as 
the distance between the superior border of the first lumbar vertebra 
(pyeloureteral junction) and the pubic symphisis (uretero-vesical junction).

Dolicomegaureter cases are problematic. In those patients the real ureteral extension 
cannot be measured directly and is bigger than topographic estimations. 
Cystourethrography exams (CUM) imply length variations due to the distance between 
patient, X-ray emission equipment and radiographic films and cannot be used to measure 
directly the ureteral length. We suggest that their length may be calculated using 
proportional formulas:

\begin{eqnarray}
{Rul \over Rptd}={ureteral.length.from.CUM \over Rptd.from.CUM}
\end{eqnarray}

\begin{eqnarray}
{Rul \over L1-symphisis.distance.in.vivo}={ureteral.length.from.CUM \over L1-symphisis.distance.from.CUM}
\end{eqnarray}

\begin{eqnarray}
Rul={ureteral.length.in.CUM \times L1-symphisis.distance.in.vivo \over L1-symphisis.distance.from.CUM}
\end{eqnarray}

\noindent Obviously, the total volume of the urinary system proximal to the bladder corresponds 
to the sum of the 2 renal pelvises to the 2 ureters.
Normal bladder capacity in childhood can be estimated from various formulae, that 
consider that the bladder grows in different paces in infants $(< 2 years-old)$ and 
older children. There are various formulae in literature. We opted here to use 
Koff's formula \cite{koff} for older children and Kaefer's formula \cite{kae} for infants, based on 
their ample use, easiness and recommendations of the International Childs Continence Society \cite{neveus} .

$Age = 2 years: Bladder volume (ml) = (age (years) + 2) . 30$
$Age < 2 years: Bladder volume  = (2 . age (years) + 2) x 30$

\noindent Those calculations allow us to obtain some important physiopathological information:

In $a=2$ years-old child $a>10\%$ proximal urinary system urine sequestration 
(corresponding to an abnormal post-micturating residue) implies that:

\noindent $(age+2)\times 30\ge (pv+uv)$

\noindent If the ureter is not a dolicomegaureter:
$(age + 2) \times 30 > 1.04 \times h \times B \times b + 3.14 \times (r/2)^2 \times (age + 10)$

\noindent Considering similar dimensions of proximal abnormally dilated urinary 
systems, urinary sequestration is proportionally more severe in infants than in 
older children, physically speaking:

\noindent In a 4 months-old infant presenting renal pelvis dimensions of $2 \times 1.5 \times 5$cm 
and a 16cm dolicomegaureter ($> 60\%$ than the normal length for his age) and 1.6 cm diameter:

\noindent $pv= 1.04 \times 2 \times 1.5 \times 5 = 15.6 ml$

\noindent $uv= 3.14 \times (1.6/2)\times 2 \times 16 = 3.14 \times 0.8 \times 0.8 \times 16 = 32.15$

\noindent Total supra-vesical urinary system volume(T) $T = 15.6 + 32.15 = 47.75$

\noindent Bladder calculated capacity(bcc) $bcc = (2 \times 0.3 + 2) \times 30 = 78 ml$

\noindent Sequestrated urine/bladder capacity(su.bc) $su.bc = 47.75/39.5 = 62\%$

\noindent In a 6 years-old male with the same pyelic dimensions and a 26 cm dolicomegaureter
($> 60\%$ than the normal length for his age) and 1.6 cm diameter:

\noindent Pelvic volume = 1.04 x 2 x 1.5 x 5 =  15.6 ml

\noindent Ureteral volume = 3.14 x (1.6/2)2 x 26 = 3.14 x 0.8 x 0.8 x 26 = 52.25

\noindent Total supra-vesical urinary system volume = 15.6 + 52.25 = 67.85

\noindent Bladder calculated capacity = (6 + 2) x 30 = 8 x 30 = 240 ml

\noindent Sequestrated urine/ bladder capacity = 67.85/240 = 28.3\%

\section{DISCUSSION:}

A non-contractile volumetric storage compartment coupled to the bladder 
augments the reservoir without a functional (contractile) correspondence. 
Physically, this means that the relationship between anatomical bladder 
volume/functional bladder volume diminishes, or that the work efficiency 
of the detrusor is reduced. 
Normal micturition begins with a neurologically mediated isovolumetric 
contraction of the detrusor, generating enough pressure to open the bladder 
sphincters. From that moment on the urine is expelled with a constant pressure 
maintained by the detrusor contraction (isobaric phase), sufficient to overcome 
the passive urethral resistance. During isovolumetric detrusor contraction, 
if the patient presents VUR or a big bladder diverticulum this low pressure 
non-contractile "reserve systems" store a fraction of the urinary volume to 
be expelled (communicating vessels principle), depending on the characteristics 
of the reservoir (volume of the proximal urinary systems or diverticula, 
presence of absence of obstructions to retrograde urine flux, position of the 
patient as it relates to gravity). Perhaps even more important, a non-contractile 
communicating system dissipate some of the energy generated by the detrusor 
contraction, causing a less efficient bladder systole. There are some direct 
consequences of this process:

\begin{enumerate}
  \item {A more intense detrusor contraction (more detrusor work) is necessary 
   to maintain a satisfactory bladder emptying or, alternatively, higher 
   bladder residua appear after each micturition. This may cause detrusor 
   secondary hipertrophy, detrusor-sphincter secondary incoordination and 
   bladder myogenic decompensation. By this mechanism, VUR may be the cause, 
   and not the effect, of bladder dysfunction.  As a matter of fact, the 
   two defects (VUR and bladder dysfunction) feedback each other, even if 
   we believe that a common embryological malformation could cause both the 
   VUR and a primary urodynamic problem, secondary to trigonal and/or neural 
   dysgenesis.}

\item{After a micturition, during bladder diastole, high tract sequestrated 
urine returns to the bladder. Mechanical fluid laws attest that liquids 
move between two communicating systems according to pressure gradients 
between the two compartments. During bladder diastole the bladder has 
the same pressure of the abdomen and gravity facilitates bladder filling 
from the pelvises and ureters in orthostatic patients. Obviously this 
may not apply to small infants that lie supine for the major part of the day.}

\item{High bladder post-micturating residua are generated from detrusor inneficiency 
and abnormal storage of urine in the ureters and renal pelvises. Depending on the 
final residual volume, bladder sensitivity and the motor efficiency of the detrusor, 
a new emptying cycle initiates immediately after the preceding one, worsening detrusor 
fatigue and progressive inefficiency. If the new emptying cycle does not occur 
(for example in neuropathic CIC-dependent bladders) or is inefficient, bladder 
urinary stasis augments the risk of UTI.}

\end{enumerate}

If the functional cystometric capacity, complacency or sensitivity to repletion 
are abnormally low those decompensation mechanisms are worsened, as the storage 
function of the bladder will be as jeopardized as its motor function. In those 
patients hydronephrosis tend to progress. Neuropathic bladders, valve bladders, 
small bladders and "reflex" bladders (typical of infants) are the most susceptible. 
Abnormal bladder function is common in male infants with severe VUR. Our 
calculations demonstrate that the dynamic implications of urinary sequestration are 
worse in infants than in older children. Most authors agree that VUR can be worsened 
or caused by "bladder disfunction", but the ideas expressed here suggest that the 
exact inverse phenomenon may happen: bladder dysfunction may be secondary to severe 
dilating long-term VUR. This is a new concept, and can be a good argument to treat 
VUR aggressively in those patients, even without UTI. These ideas can explain the 
cure of bladder dysfunction after VUR endoscopic or surgical treatment reported 
recently from some authors \cite{lack,musq}. 
Yeung and cols \cite{yeu}, studying the urodynamic patterns of normal and severe VUR infant 
males describe $57\%$ of urodynamic abnormalities in the index patients 
(none in the normal infants), typically high post-micturating residua, high bladder 
pressures and hyper active bladder pattern. Podesta et al \cite{podesta} corroborate their 
findings while studying infants with VUR, but do not report bladder dyastole problems.  
Sillen et al \cite{sillen} present 2 types of urodynamic abnormalities in VUR infant patients, 
emphasizing that low capacity hyper dynamic bladders are characteristic of males. 
Infants frequently present an immature bladder-sphincteric coordination, but normal 
infants do not present high urinary residua. In another article, Yeung et al \cite{yeu1} document 
abnormally thick bladder walls in infants with VUR.  
Our calculations evaluated more exactly the high tract sequestration, and compared it 
to bladder capacity. Our results may suggest therapeutic manoeuvers (pharmacological 
or surgical bladder augmentation, timed voiding, VUR active treatment, ureterostomy, 
vesicostomy, nocturnal continuous bladder catheterization) to be indicated in some 
patients before severe consequences of reflux appear. Sequestration $> 20\%$ of the bladder 
capacity probably means a high urodynamic risk.: as a matter of fact, International 
Children's Continence Society has recently suggested that $> 20 ml$ residua in children 
is abnormal, and in adults $> 10\%$ post-voiding residua are considered dangerous \cite{kae} There 
is at least one manuscript attesting the cure of serious urodynamic problems after VUR 
treatment, without any other therapeutic measures, in a patient presenting with $> 50\%$ 
bladder capacity urinary sequestration \cite{musq}. Mialdea and colleagues have described a 
"posterior urethral valve-like syndrome", presenting VUR -associated to uncoordinated 
fetal voiding in male infants and suggest that a vesicostomy can simultaneously cure 
VUR and urodynamic abnormalities in 2/7 babies\cite{mial}
The usage of our formulae can be complementary to VUD, and corroborate its results 
The logistical problems with the exam, especially concerning neurologically normal 
infants, radiation exposure, high cost of the equipment and low availability of experts, 
mainly in poor countries or non-reference smaller hospitals \cite{yeu1,mcguire}, render the direct 
calculations attractive, at least as an initial triage of patients who should be 
referenced to surgical treatment or evaluation.
This mathematical model, though logical, needs to be tested clinically. Ideally the 
calculations should be confronted to direct measurements taken with image exams 
$(3d CT or MRI)$, as in vivo direct measurements cannot be done. Limited clinical 
indications of those exams in VUR patients, radiation exposure and the needing for 
sedation create ethical and practical problems to obtain those data. Another problem is 
that there are some different formulae proposed to calculate normal bladder capacity in 
children. We opted to use a formula for infants and another for $children = 2 years-old$, 
as various authors have already demonstrated the logarithmic bladder growth in childhood, 
with an accelerated period during infancy. Bael and cols suggested that bladder capacities 
are similar before and after VUR cure, not relating to VUR grade or laterality, and 
proved that refluxing volumes correlate primarily with reflux grades (and, supposedly, 
ureteral and pyelic dilation)\cite{bael}. Kaefer's formula was suggested after a clinical study 
with $> 2000$ children \cite{kae} and Koff's formula \cite{koff} has been recently adopted by the International 
Children's Continence Society \cite{neveus}.
 
\section{CONCLUSION:}

Sequestration of urine in the ureter and pelvis can be estimated from mathematical formulae 
in patients with VUR. The values used derive from ultrasound examinations, CUM and 
topographical anatomical references. VUR can determine directly urodynamic problems. 
Implications of urine sequestration in the proximal urinary system are worse in infants 
than in older children.

\end{document}